\documentclass[conference]{IEEEtran}

\usepackage{graphicx}
\DeclareGraphicsExtensions{.pdf,.jpeg,.png}

\begin{document}
%
\title{Stack up your chips: Betting on 3D integration to augment Moore’s Law scaling}

\author{\IEEEauthorblockN{Saurabh Sinha, Xiaoqing Xu, Mudit Bhargava, Shidhartha Das, Brian Cline and Greg Yeric}
\IEEEauthorblockA{Arm Research,
5707 Southwest Parkway, Austin, Texas 78735\\
Email: saurabh.sinha@arm.com}}


%

\IEEEspecialpapernotice{(Invited Paper)}

\maketitle

\begin{abstract}

3D integration, i.e. stacking of integrated circuit layers using parallel or sequential processing is gaining rapid industry adoption with the slowdown of Moore’s Law scaling. 3D stacking promises potential gains in performance, power and cost but the actual magnitude of gains varies depending on end-application, technology choices and design. In this talk, we will discuss some key challenges associated with 3D design and how design-for-3D will require us to break traditional silos of micro-architecture, circuit/physical design and manufacturing technology and work across abstractions to enable the gains promised by 3D technologies.
\vspace{6pt}
\hrule\relax
\vspace{6pt}
This is an accepted version of the IEEE published article
presented at IEEE S3S Conference 2019, San Jose, USA / http://s3sconference.org. \\
\\
\copyright  2020 IEEE. Personal use of this material is permitted.
Permission from IEEE must be obtained for all other uses, in any
current or future media, including reprinting/republishing this
material for advertising or promotional purposes, creating new
collective works, for resale or redistribution to servers or
lists, or reuse of any copyrighted component of this work in
other works.


\end{abstract}


%
\IEEEpeerreviewmaketitle

\section{Introduction}
The semiconductor industry has achieved unprecedented growth in the last six decades owing to its incessant drive to fulfill the prophecy of Moore's Law scaling \cite{moore:1965}. Moore's law continued to provide value to the semiconductor industry as cost per transistor reduced with shrinking feature sizes. However, as we hit physical limits of transistor scaling and increasing cost of  lithography and patterning, the industry is transitioning to design-technology and system-technology co-optimization (STCO) paradigms where added value is achieved through heterogenous integration of different technologies targeted towards specific end-applications \cite{yeric:2015}. 2.5D and 3D stacking techniques are key enablers of this new paradigm.

3D integration is a wide term encompassing technologies that enable vertical integration of more than one layer of active transistors and interconnects with the goal of increasing compute density. 
Integrated circuit (IC) designs with natural redundancy and regularity in 2D can be extended or stacked in the 3rd dimension with relative ease. CMOS image sensors \cite{fontaine:2019}, DRAM memories \cite{jun:2017}, and NAND Flash memories \cite{venkatesan:2018}, are all examples of this type of IC, and these products have already adopted 3D integration and achieved success in high-volume market adoption.

However, adoption of 3D stacking for logic applications has been limited to advanced packaging techniques. Here functionally complete chips, commonly referred to as chip-lets, are stacked using package bumping technologies. The stacking configuration could be 2.5D, wherein, chip-lets are assembled in 2D but interconnected through an underlying substrate (e.g., Silicon interposer) or redistribution layer (RDL), e.g., fan-out RDL. Alternatively, the stacking configuration could be 3D, e.g., package-on-package (PoP) wherein DRAM packaged dies are stacked on ASIC die \cite{tseng:2016} or two or more compute dies stacked using through-silicon-via (TSV) and micro-bump technology \cite{foveros:2019}. A discussion of advanced 3D packaging using bumping technologies is out of scope of this paper. 

The trajectory of current adoption of 3D stacking technologies points towards finer-pitch 3D connectivity in the form of die-stacking or sequential 3D integration, which, we refer to as high-density 3D integration.  High-density 3D integration techniques open the possibility of designing systems where functional units are partitioned and co-designed across separate 3D stacked tiers. The advantages of such 3D integration is multi-fold:

\vspace{-2.5mm}
\begin{itemize}
	\setlength\itemsep{0em}
\item Systems can utilize the 3rd dimension to bring functional blocks closer, reducing interconnect delay and power.
\item Large die SoCs can be partitioned into smaller dies, improving yield and hence reducing cost.
\item Dies from different process nodes or technologies (e.g., non-volatile memory) can be integrated together enabling heterogenous integration and enables more flexible product migration to advanced nodes further reducing cost.
\end{itemize}
\vspace{-2.5mm}

However one of the primary challenges/opportunities that will be required to fully access the above advantages will be the re-design of design architecture to take advantage of computing and memory density that is different than what we have come to know in decades of 2D integration. The rest of the paper provides an overview of high-density 3D stacking technologies, state-of-the-art physical design studies and associated challenges. The paper concludes with a motivation for 3D-aware architecture exploration that breaks the traditional silos of the semiconductor design ecosystem.

\section{High-density 3D technologies}
\begin{figure}[!t]
\centering
\includegraphics[width=0.8\columnwidth]{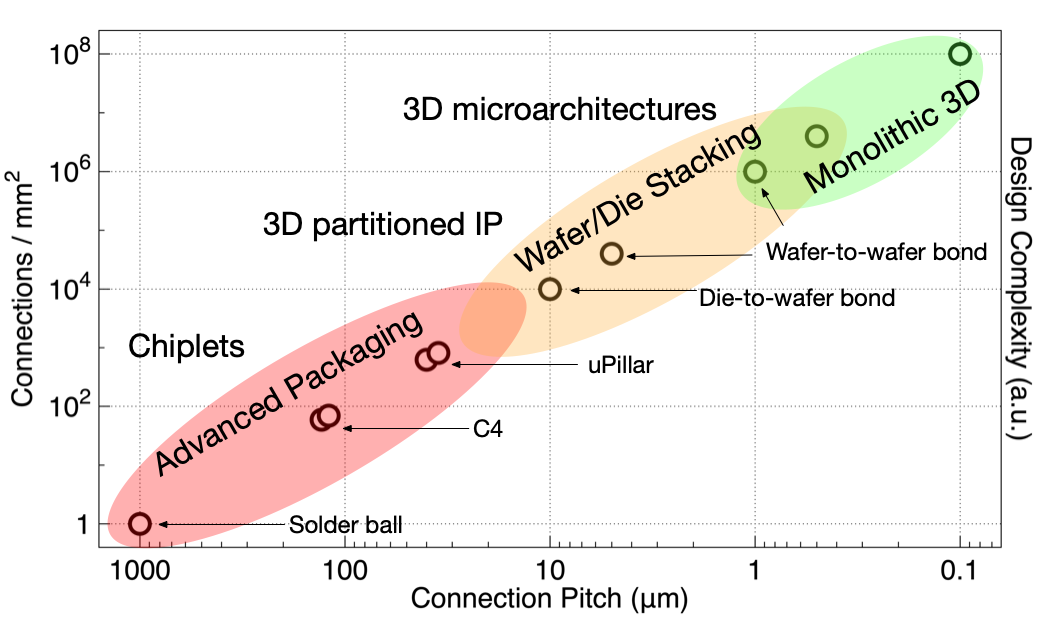}
\caption{Comparison of connection pitch and density of different 3D technologies ranging from advanced packaging, die-stacking and monolithic 3D.}
\label{fig1_3d}
\end{figure}

Current adoption of 3D stacking is mainly in the packaging domain and 3D connection density is limited by bump pitches at approximately 40 $\mu m$. However, wafer-level and die-level stacking technologies such as hybrid-bonding allow precision alignment of wafers resulting in 3D connection pitches of 10 $\mu m$ or less \cite{chen:2019}. At these 3D connection pitches, SoC functional unit partitioning becomes feasible. The 3D integration roadmap is shown in Fig.\ref{fig1_3d} as a plot of connection pitch versus connection density which highlights the orders of magnitude higher 3D connections that are feasible as we transition from package level bumping technologies to hybrid wafer bonding techniques. 

Another flavor of high-density 3D is monolithic or sequential 3D integration where two or more active device layers and interconnects are sequentially processed using standard lithography tools. The 3D connection pitch is limited by the alignment of lithography stepper tools, enabling pitches down sub 100nm pitch, i.e., metal via pitches at advanced process nodes. However, this technology faces challenges with incompatibility of BEOL and FEOL processing temperatures for silicon based transistors \cite{batude:2015}. Alternative approaches of using materials and devices that do not require high temperature processing such as carbon nanotube field effect transistors (CNFET) and resistive non-volatile memories (RRAM) have been proposed for monolithic 3D integration  \cite{aly:2019}. These technologies have seen slow but steady progress in experimental demonstrations \cite{hills:2019}.

\section{3D Cost-saving}
Any semiconductor technology promising to augment Moore's Law scaling will be required to pass the litmus test of cost scaling. 3D die-stacking technologies achieve this in a manner similar to 2.5D chiplet approach, i.e., implementing the functionality of a large monolithic die in smaller dies interconnected in 3D. Compared to 2.5D chiplet approach, 3D die-stacking can achieve significantly higher connectivity and lower latencies, hence, improving performance and power translating to added value. 

\begin{figure}[!t]
\centering
\includegraphics[width=0.85\columnwidth]{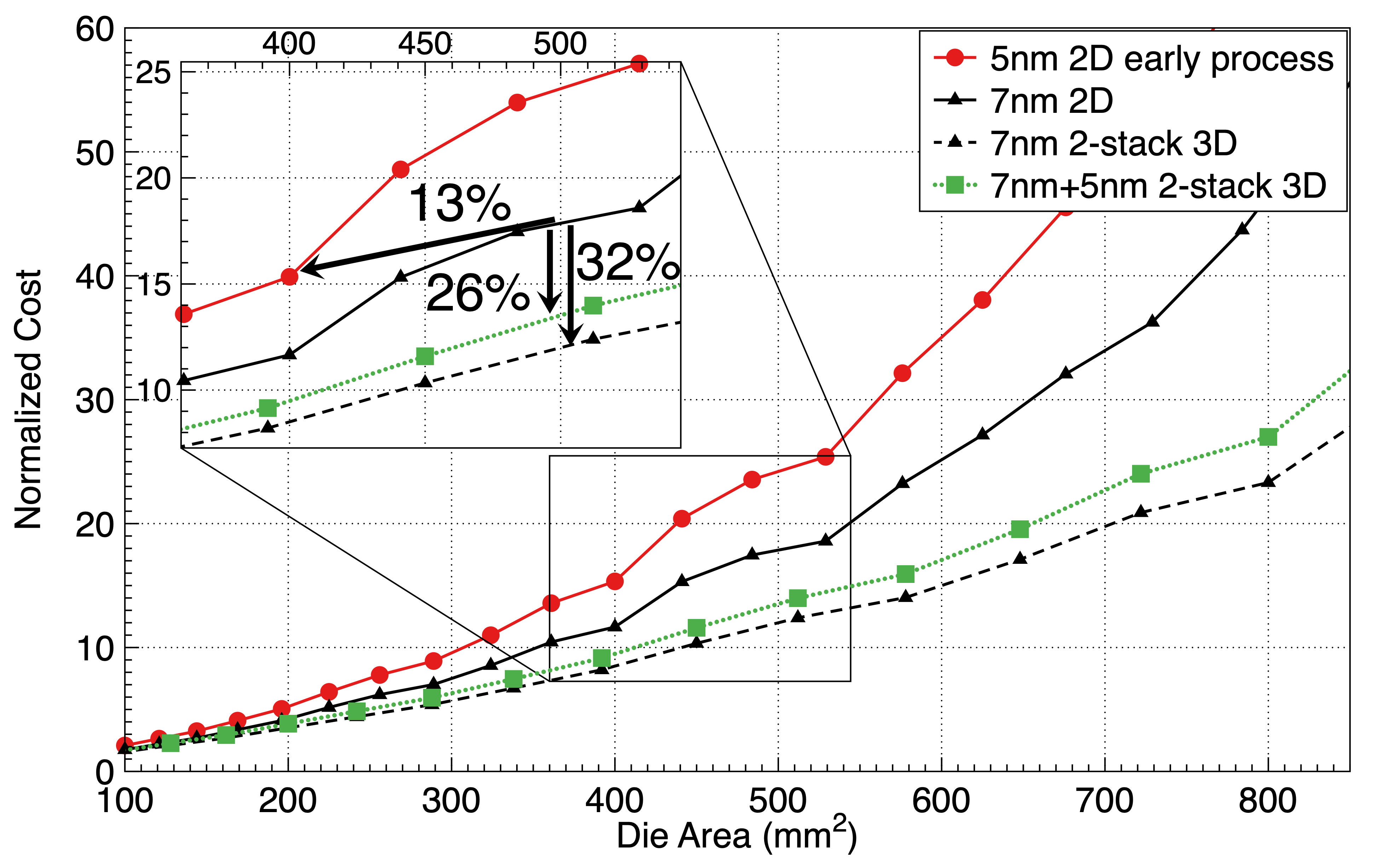}
\caption{Die cost vs. total die area for 2D and 3D stacked designs. Early ramp of 5nm assumes D$_0$=0.2 and mature 7nm process D$_0$=0.15.}
\label{fig2_cost}
\end{figure}

The value of 3D versus 2D is dependent on die size and die size is also dependent on technology node. The trade-off here is the time to market and risk-reduction of getting to market with early N-node 3D solution vs. an N+1 node 2D solution. Fig. \ref{fig2_cost} models a scenario where total die-area vs. die-cost is plotted for an early process ramp of representative 5nm technology and compared to a relatively mature 7nm process technology. Due to higher costs and worse defect densities in early ramp, 5nm die-cost for the same area is higher compared to 7nm.  The different arrows show cost trade-offs of scaling to a 5nm process versus implementing a 3D solution in 7nm or implementing a heterogenous 3D system comprising of a mix of 5nm and 7nm dies, targeting an example area of 500 $mm^2$, representative of multi-core high performance system. A conventional technology shrink gives 13\% cost reduction while a heteregenous 3D solution of a 5nm and 7nm stacked die doubles the cost benefit to 26\% lower die-cost and a 3D solution at 7nm gives a 32\% lower cost. Breaking an SoC into logic and memory layers where memory layers are repairable is another of the many possible embodiments of 3D that could have varying degrees of benefit per product.

For 3D cost savings to be realized as modeled, test for known-good-die (KGD) is a requirement prior to 3D assembly, hence is only applicable to die-to-wafer stacking scenarios. Since high-density 3D can have connections at sub-10$\mu m$ pitches, direct probing of every 3D connection priori to assembly is non-trivial and can be expensive. Novel Design-for-test (DFT) techniques for 3D stacking need to be developed that allow testing each die for 'goodness' prior to assembly. There are active efforts in standardizing DFT methodologies for 3D in the form of IEEE P1838 standard \cite{marinissen:2016}. 

Assuming design and test challenges are addressed, additional cost savings could be achieved through 3D STCO instead of trying to reduce cost per transistor through large die splitting. As an example, a 3D optimized N-core system could potentially perform equally to a 2D M-core system (where N$<$M) due to improved bandwidth and connectivity. 

Detailed cost modeling of monolithic 3D designs has been presented in \cite{gitlin:2016} and \cite{ku:2016}. The primary yield improvement in monolithic 3D comes from the fact that the critical area for defect densities can be reduced by approximately 2X in monolithic 3D wafer processing. Considering different scenarios, these works have found that monolithic 3D can enable cost savings compared to 2D designs, especially for large die areas.

\section{3D physical design}

\begin{figure}[!t]
\centering
\includegraphics[width=0.8\columnwidth]{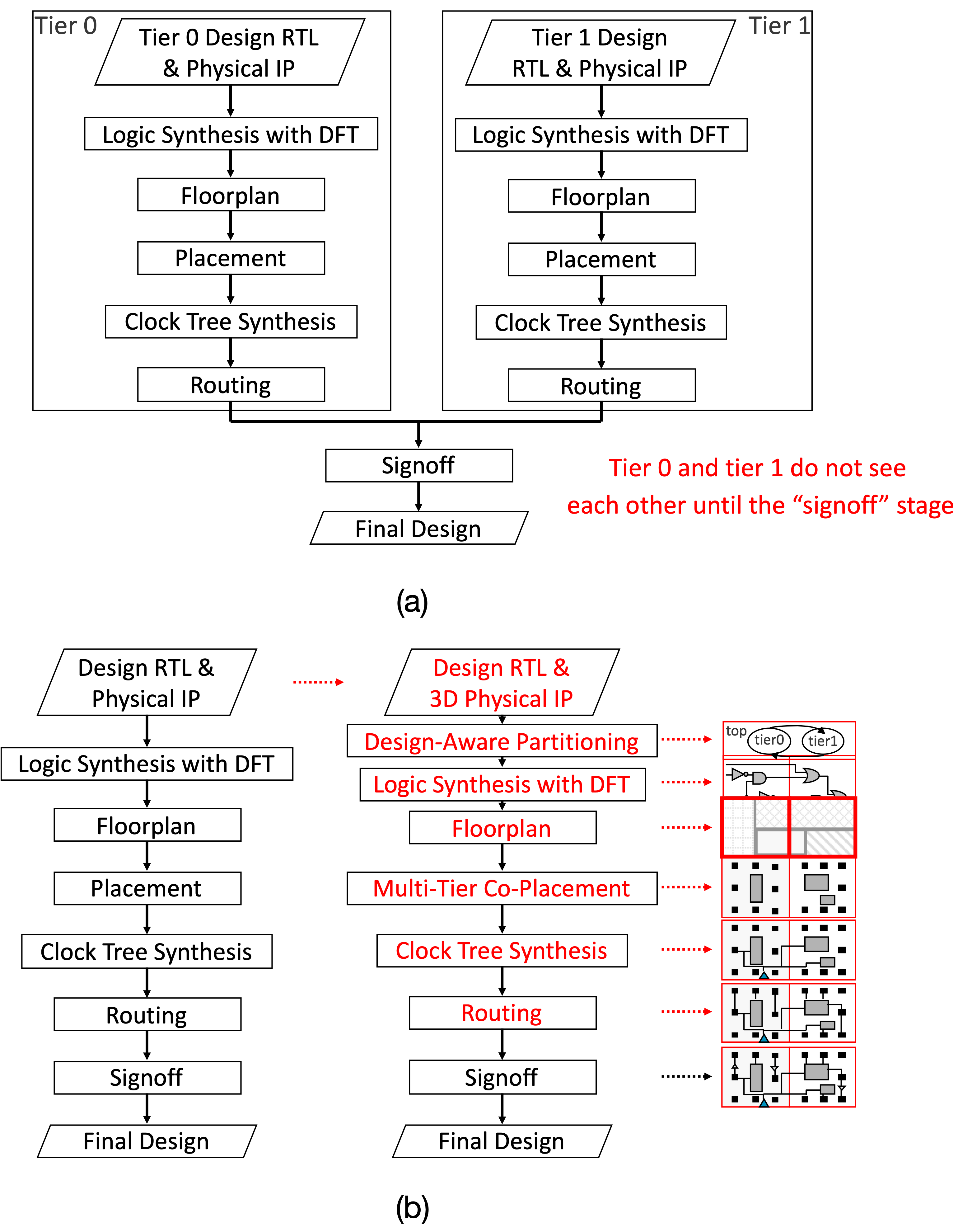}
\caption{(a) State-of-the-art commercial 3D design flow (b) 3D design flow using 2D EDA tools.}
\label{fig_3dflow}
\end{figure}

3D design has been explored extensively in the past few decades based on through-silicon-via (TSV) technology assumptions. High-density 3D design explorations that enable design partitioning at a block or gate level have been challenging, mainly because of the lack of EDA tools to implement such designs. Fig. \ref{fig_3dflow} (a) shows current state-of-the-art 3D physical design flow supported by EDA tools today. Today's 3D-IC designs are predicated on the assumption that functionally complete systems would be stacked in 3D. Hence, current tools do not support any automated 3D partitioning or cross-tier placement, timing or routing optimizations. Each 3D tier is separately designed and optimized and cross-tier connections are only analyzed and verified during the final 'sign-off' stage. 

\begin{figure}[!t]
\centering
\includegraphics[width=0.75\columnwidth]{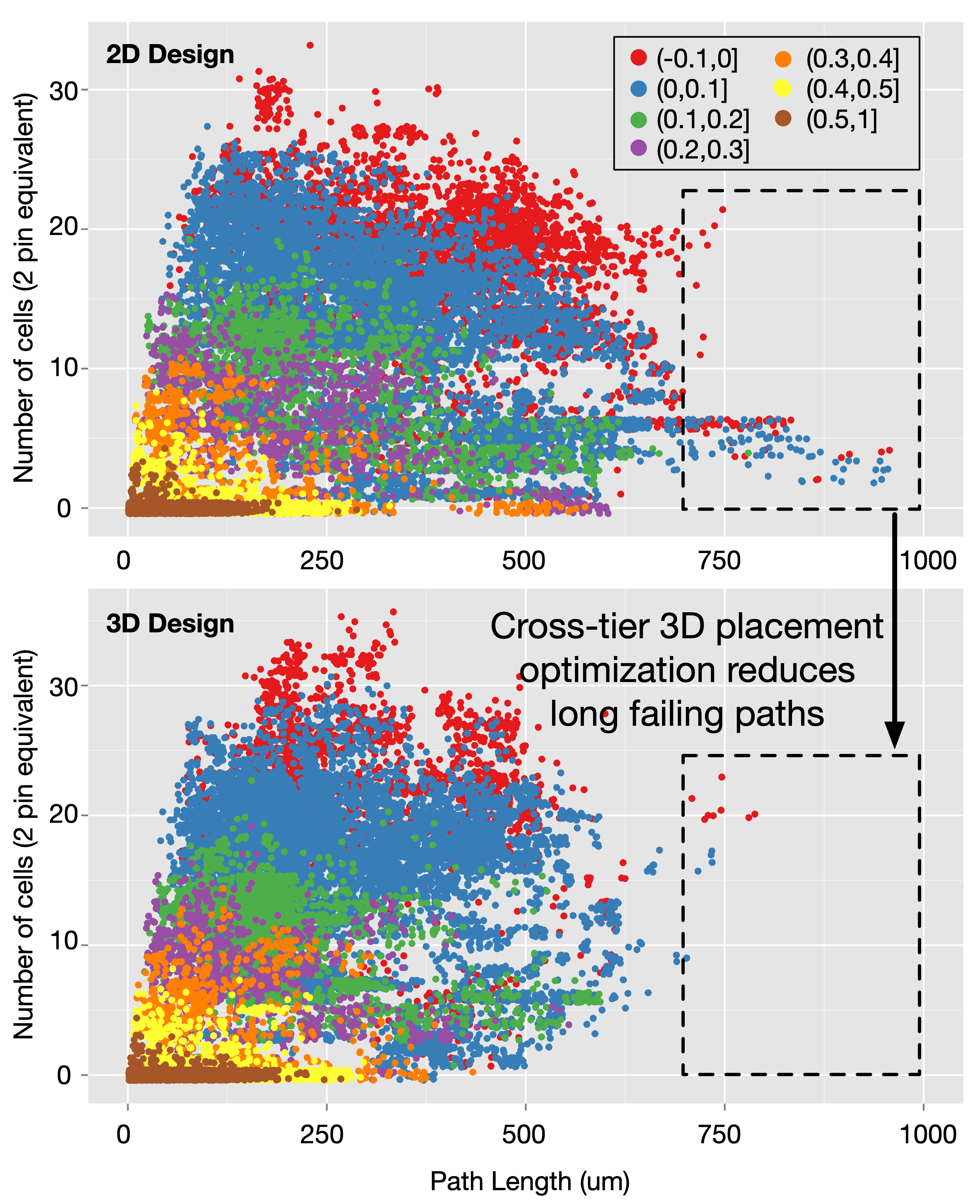}
\caption{Path complexity versus path length for varying timing slack for 2D and 3D implementation of an Arm Cortex-A CPU. Cross-tier 3D optimization reduces long timing critical paths in the design}
\label{fig_pathstats}
\end{figure}

A 3D-aware EDA tool, especially supporting high-density 3D technologies, would enable partitioning and optimized synthesis, floorplanning, placement, clock tree synthesis and routing of 3D tiers inherently in the design flow. Since these capabilities do not exist today, significant research efforts have been made to enable 3D physical implementation using 2D EDA tools \cite{panth:2017}, \cite{chang:2016}. In a recent paper \cite{xu:2019}, we presented our efforts on co-optimization of gate placement across 3D tiers using commercial EDA tools, for a 3D-partitioned Arm Cortex-A  microprocessor. The important steps of the flow and how it differs from conventional 2D design methodology is described in Fig. \ref{fig_3dflow} (b). Multi-tier co-placement utilizes commercial EDA placement optimization engine mimicking the behavior of a 3D-aware placement engine. 


Results shown in Fig. \ref{fig_pathstats} highlight the efficacy of the multi-tier co-optimization flow. In this plot, path-length versus number of cells in a path are plotted color-coded by timing slack. Path-length refers to the summation of pin-to-pin half-parameter wirelength among all net connections in the a design timing path. Number of cells denote the enumeration of two-pin equivalent gates in the timing path. For the 3D design, multi-tier co-placement efficiently places logic blocks in close proximity to each other in the 3rd dimension and is able to significantly reduce the number of long failing timing critical path. Additionally the overall path-length distribution is tighter compared to the 2D case as well.  This structural improvement in the design directly translates to performance improvement of up to 12\% or power reduction of up to 40\% \cite{xu:2019}, approaching that of a modern Moore's law process node highlighting the importance of 3D-aware tool flows.

\begin{figure}[!t]
\centering
\includegraphics[width=0.8\columnwidth]{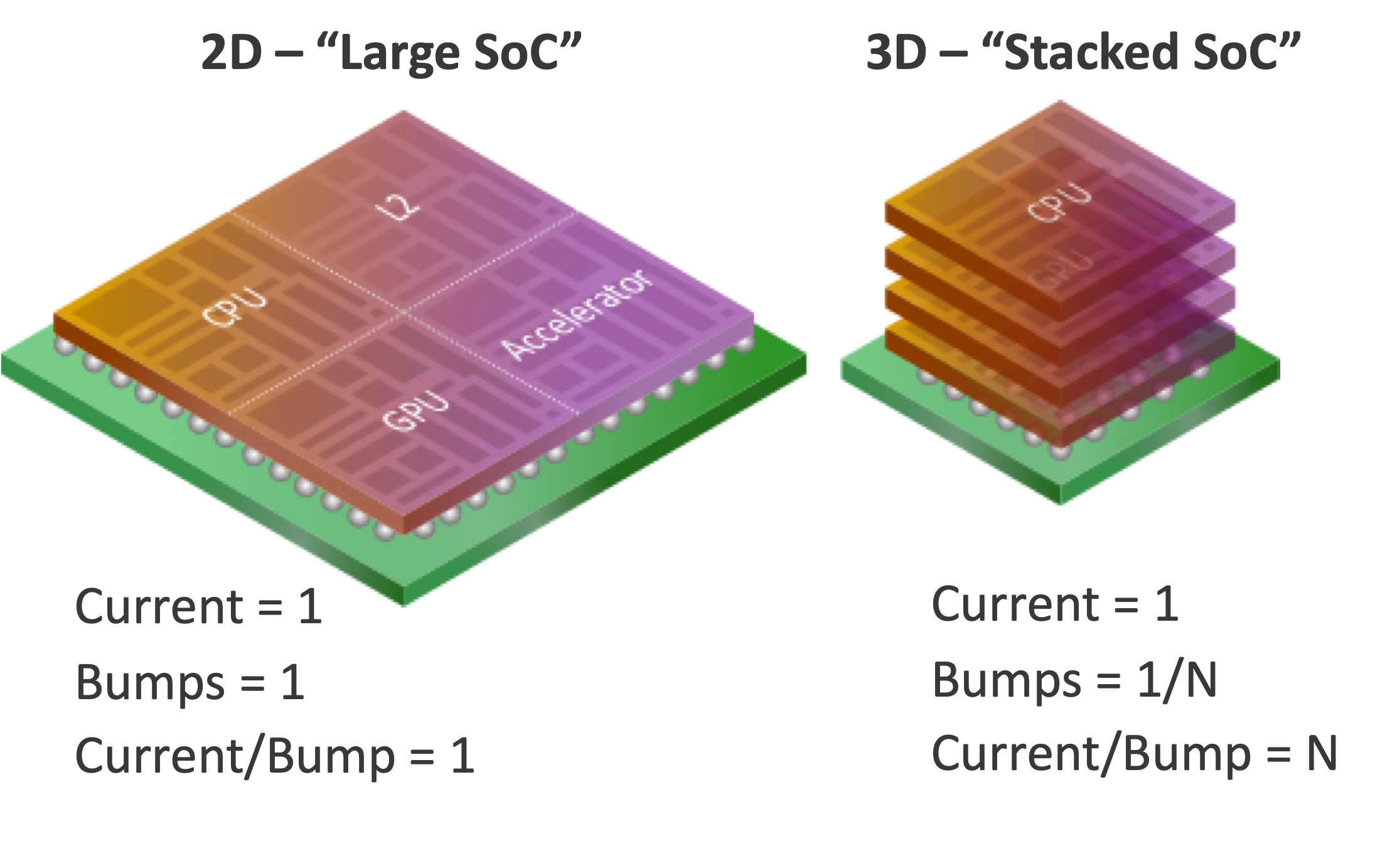}
\caption{2D vs. 3D power delivery. }
\label{fig_pdn}
\end{figure}

A key challenge in realizing 3D design implementations is designing a robust power delivery network and managing thermal dissipation. For the same design implemented in 2D versus 2-stack 3D, the 3D design occupies a smaller 2D footprint, potentially 50\% of the original design. However, the 3D design requires similar power drawn through a smaller number of package bumps in the reduced footprint increasing the current drawn per bump, as described in Fig. \ref{fig_pdn}. This directly translates to higher power density as well. These constraints require careful floor-planning of the 3D tiers to avoid power and thermal hotspots and a robust power delivery network design. It is possible that 3D systems may need more expensive packaging and cooling solutions to offset the power density increase, partially offsetting the underlying advantages of 3D stacking. Numerous solutions have been proposed to mitigate power delivery and thermal challenges in 3D designs \cite{scheurmann:2016}, \cite{chang:2019} and this is an area of active research.

\begin{figure}[!t]
\centering
\includegraphics[width=0.9\columnwidth]{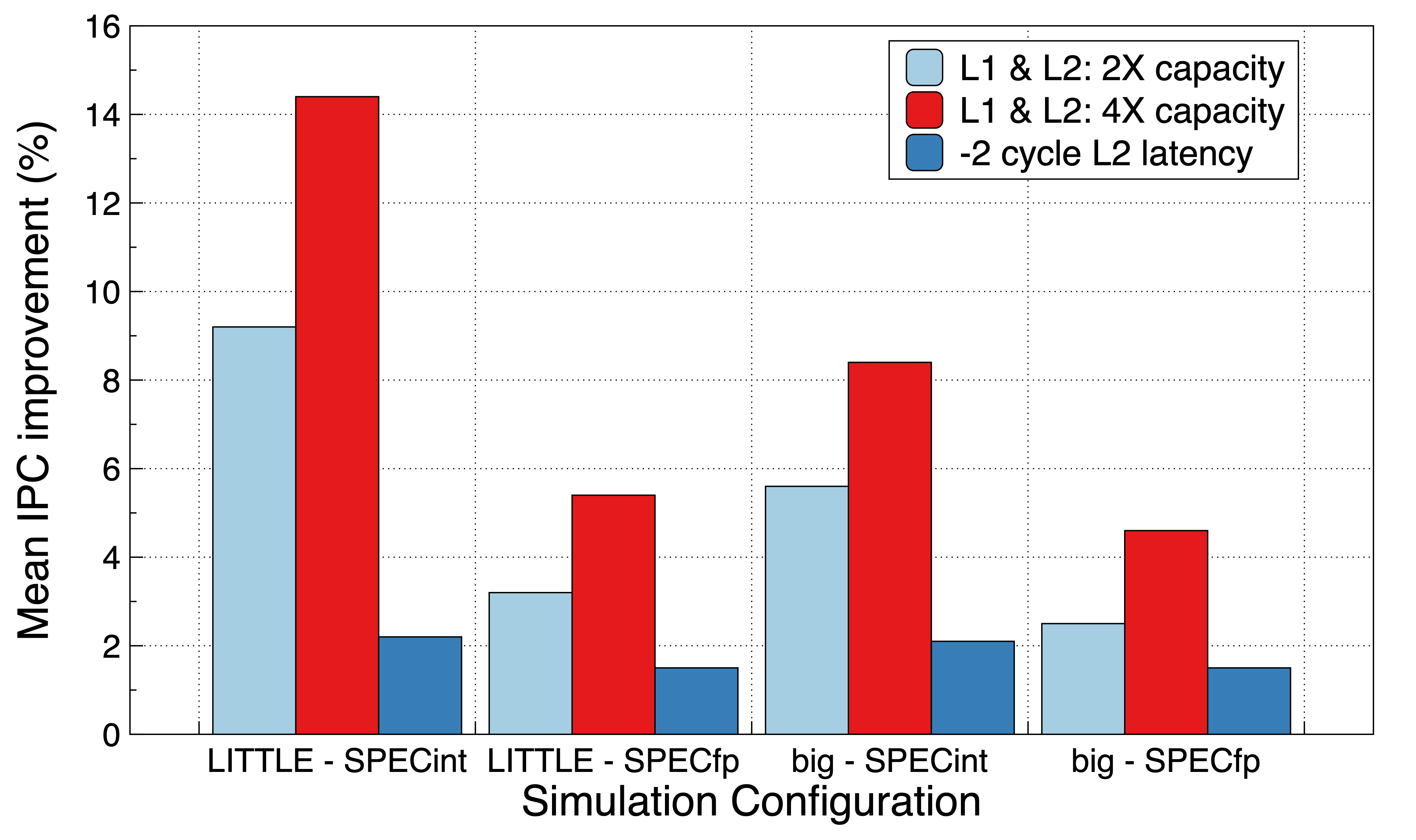}
\caption{IPC improvement of 64-bit Arm big and LITTLE CPU cores with 2X and 4X L1 and L2 capacity compared to the default configurations using gem5 simulations. }
\label{fig_gem5}
\end{figure}

\section{3D Architecture}

High density 3D poses the question whether we can fundamentally re-think design micro-architecture to take advantage of 3D stacked tiers. An area of focus for 3D micro-architecture research has been the goal of breaking the von-Neumann bottleneck, i.e., bringing larger capacity high-bandwidth memory closer to compute. Fig. \ref{fig_gem5} shows a plot of mean IPC (instructions per clock) improvement on running the SPEC benchmark suite on an Arm big and LITTLE CPU design in gem5 \cite{binkert:2011} with larger capacity and lower latency L1 and L2 caches. Significant IPC benefits can be seen by having larger capacity low latency memory access for general purpose CPUs. This concept has been extended to large-scale systems, where solutions of stacking DRAM dies over CMOS logic compute chips \cite{carlson:2011} and 3D network-on-chip (NoC) systems \cite{akgun:2019} have been proposed. A recent work proposes a monolithic 3D solution integrating compute, caches and random-access memories which does not require any off-chip memory access, essentially enabling orders of magnitude higher energy efficiency \cite{aly:2019}. Besides addressing the logic-memory bottleneck, \cite{gopireddy:2019} presented a 3D micro-architecture study of designing vertical processors using monolithic 3D technology, wherein all critical stages of a superscalar out-of-order CPU are partitioned in 3D tiers achieved significant performance improvement at lower energy dissipation. 

These works point to significant gains possible with high-density 3D stacking technologies. The magnitude of gains are work-load dependent (compute-bound versus memory-bound) and whether 3D integration effectively relieves existing 2D bottlenecks. Conventionally system architecture and micro-architectural explorations abstract out physical details in favor for cycle-accurate design behavior. This abstraction has worked well in the era of traditional Moore's Law scaling. However, this approach makes it challenging to assess realizable gains for new 3D stacked architectures since the underlying 3D technology and physical design constraints have a significant impact on achievable gains. Designing next generation high-performance general-purpose computing systems in 3D requires extensive effort and co-optimization of system and CPU architectural exploration in the context of physical design.

%


\section{Conclusions}
This paper presents an overview of high-density 3D integration technologies and its potential to improve performance, power, and cost, essentially augmenting Moore's Law scaling. 3D-optimized architectures could potentially enable higher gains. There is industry-wide effort to address 3D manufacturing and design challenges in the form of 3D-aware EDA tools, robust power delivery, thermals, and test of known-good-die. As high-density 3D manufacturing technologies and physical design methodologies mature, it is time to revisit 3D-optimized architecture research with strong cross-abstraction collaboration between technologists, circuit designers and computer architects.

\bibliographystyle{unsrt2authabbrvpp}
\bibliography{ref1}
%



\end{document}